\documentclass[aps,prl,onecolumn,showpacs,superscriptaddress]{revtex4-2}  

\usepackage{chemformula} 
\usepackage[T1]{fontenc} 
\usepackage{bm}
\usepackage[normalem]{ulem}
\usepackage[T1]{fontenc}
\usepackage[latin9]{inputenc}
\usepackage{color}
\usepackage{units}
\usepackage{amssymb}
\usepackage{graphicx}
\usepackage{esint}
\usepackage{bm}
\usepackage{natbib}
\usepackage{ulem}
\usepackage[colorlinks=true, citecolor=red]{hyperref}

\begin{document}
	
	\title{Multifractal conductance fluctuations in high-mobility graphene in the Integer Quantum Hall regime}

	\author{Kazi Rafsanjani Amin}
	\affiliation{Department of Physics, Indian Institute of Science, Bangalore, Karnataka, India 560012}
	\affiliation{Current address: Department of Microtechnology and Nanoscience, Chalmers University of Technology, 412 96 Gothenburg, Sweden}
	
	\author{Ramya Nagarajan}
	\affiliation{Department of Physics, Indian Institute of Science, Bangalore, Karnataka, India 560012}
	\author{Rahul Pandit}
	\affiliation{Department of Physics, Indian Institute of Science, Bangalore, Karnataka, India 560012}
	\author{Aveek Bid}
	\email{aveek@iisc.ac.in}
	\affiliation{Department of Physics, Indian Institute of Science, Bangalore, Karnataka, India 560012}

	\begin{abstract}
		
We present the first experimental evidence for the multifractality of a
transport property at a topological phase transition.  In
particular, we show that conductance fluctuations display
multifractality at the integer-quantum-Hall $\nu=1
\longleftrightarrow \nu=2$ plateau-to-plateau transition in a
high-mobility mesoscopic graphene device. We establish that to
observe this  multifractality, it is crucial to work with very
high-mobility devices with a well-defined critical point. This
multifractality gets rapidly suppressed as the chemical
potential moves away from these critical points.  Our
combination of multifractal analysis with state-of-the-art
transport measurements at a topological phase transition
provides a novel method for probing such phase transitions in
mesoscopic devices.  
		
	\end{abstract}
	
	\maketitle
	
\begin {figure}[!t]
	\begin{center}
		\includegraphics[width=0.75\columnwidth]{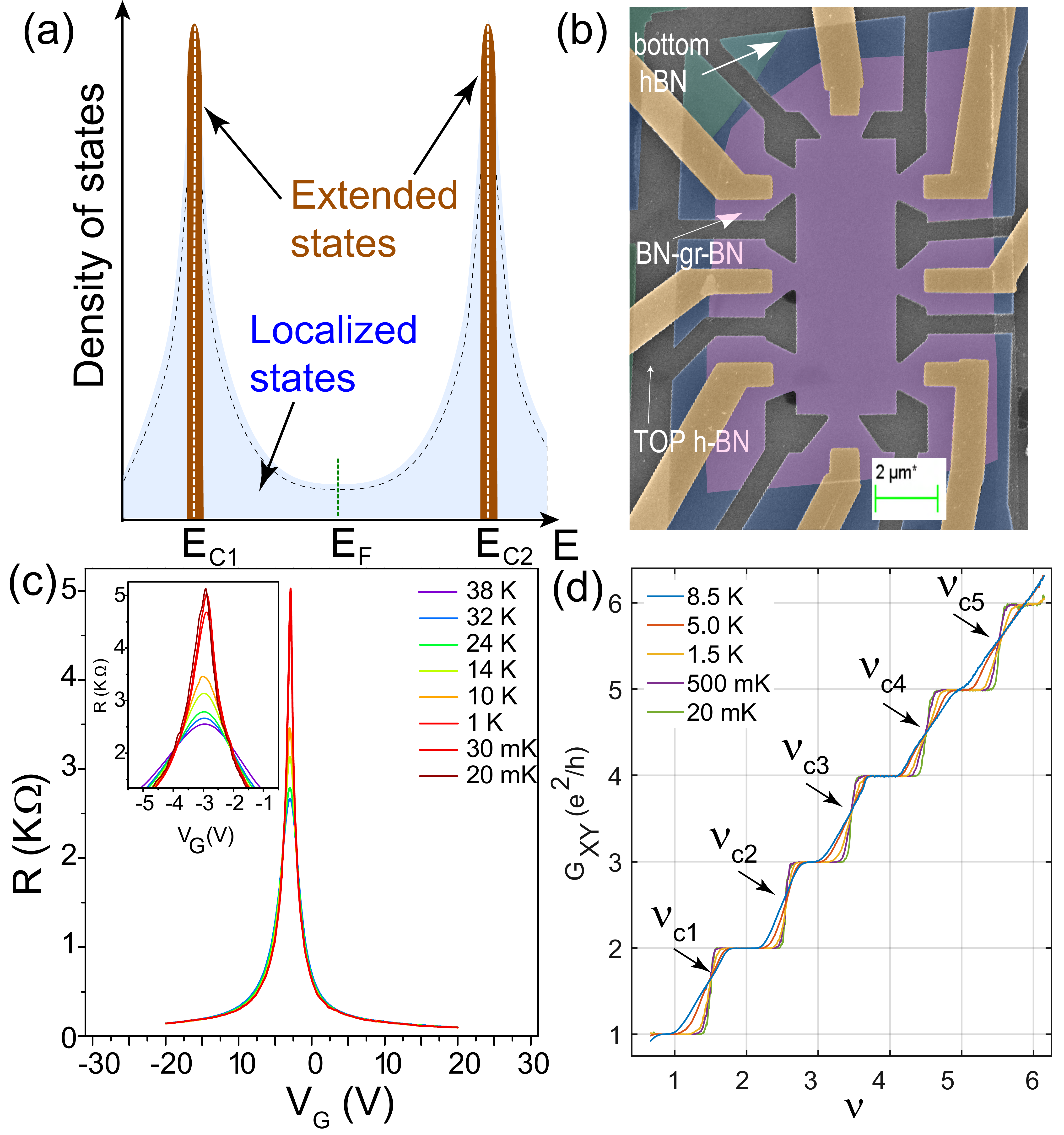}
		
\caption{(a) A schematic diagram illustrating the dependence of the density of
states (DOS) of the 2DEG on energy in the IQH regime, showing brown regions 
with extended states, at the centers of the disorder-broadened Landau levels,
separated by light-blue regions with localized
states. Plateaux in the Hall conductance are
observed when the Fermi energy $E_F$ (green dashed line) lies
deep within the band of localized states. The quantum phase
transitions between the localized and extended states occur at
the mobility edge around the critical energies
$\mathrm{E_{C1}}$ and $\mathrm{E_{C2}}$. (b) A false-color SEM image
of our device 1DC8 (see text).  (c) Plots showing the
dependence of   $R$ on $V_G$ at different temperatures. The
measurements were carried out at $B=0$~T. The inset shows a
zoomed-in plot near the Dirac point. (d) Plots of $G_{XY}$
versus the filling factor $\nu$ measured at different
temperatures and magnetic field $B=16$~T. The arrows mark the
positions of the IQH critical points at which the 
localization-delocalization transitions occur.
\label{fig:sem-rvg}}
		
	\end{center}
	\end {figure}

	
Since its discovery, the integer quantum Hall (IQH) effect, a continuous
quantum-phase transition in a two-dimensional electron gas
(2DEG)\cite{girvin1987quantum}, has provided us with a paradigm for topological
phase transitions. In the presence of a large magnetic field $B$, applied
perpendicular to the surface, the density of states (DOS) of a non-interacting
2DEG breaks into discrete, quantized Landau levels. Disorder broadens these
degenerate Landau levels into bands of extended states that are separated by
localized states. When the Fermi level $\mathrm {E_F}$, which we can tune by
changing either $B$ or the charge-carrier density $n$, lies in the part of the
spectrum with localized states (cf. Fig.~\ref{fig:sem-rvg}(a)),  the Hall
conductance $G_{XY}$ is quantized in units of $e^2/h$, and the transverse
conductance $G_{XX}$ becomes vanishingly small, with $G_{XX}=0$ at temperature
$T=0$~\cite{klitzing1980new}.  In this regime, transport takes place through
chiral edge modes, whose number is dictated by the topological Chern number of
the
system~\cite{avron1994charge,thouless1982quantized,bellissard1994noncommutative,avron2003topological}.
If, by contrast, $\mathrm {E_F}$ lies in the range of energies at the center of
the Landau levels with extended states, transport proceeds through the bulk
with $G_{XX} \neq 0 $ and a non-quantized $G_{XY}$. The
localization-delocalization transition occurs in a 2DEG in the IQH regime as
the system crosses the mobility edge, which separates localized and extended
states~\cite{huckestein1995scaling, li2009scaling, chalker1999integer,
chalker1988scaling}. The eigenstates at the mobility edge are \textit{critical}
and different from both localized and extended
states~\cite{huckestein1995scaling}. As the Landau-level filling factor $\nu$
approaches its critical value, the  localization length $\xi$ diverges
algebraically as $\xi \propto |\nu-\nu_C|^{-\gamma}$. Theoretical studies have
shown that observables like the distribution of the local density $|\psi ({\bf
r})|^2$~\cite{huckestein1995scaling,huckestein1990one} or the equilibrium
current density $|j({\bf r})|^2$~\cite{huckestein1992multifractal} display
multifractality of the density fluctuations that leads to anomalous
diffusion~\cite{chalker1988scaling} and, consequently, a power-law decay of the
density correlations, a slow decay of temporal wave-packet
auto-correlations~\cite{huckestein1994relation} and, most significantly for our
purpose, multifractal conductance fluctuations~\cite{amin2018exotic,
benenti2001quantum,casati2000fractal,facchini2007multifractal, doi:10.1116/6.0001337}. 

In this Letter, we present the first experimental evidence for the
multifractality of a transport property at a topological phase
transition~\cite{avron2003topological,thouless1982quantized,bellissard1994noncommutative,avron1994charge}. In particular, we show that, in high-mobility-graphene at the first IQH plateau-to-plateau transition, the
conductance shows multifractal fluctuations as a function of $\nu$.
	
Multifractality was initially introduced to characterize the statistical
properties of fluid turbulence~\cite{tsinober1996turbulence,ghil1985turbulence}
and thereafter studied, not only in turbulent
flows~\cite{benzi1984multifractal, pandit2009statistical, pal2016binary}, but
also in a variety of fields like the analysis of DNA sequences
~\cite{buldyrev1998analysis}, atmospheric science ~\cite{zeng2016scaling,
bhattacharjee2020anisotropy}, econophysics~\cite{mantegnaintroduction},
heartbeat dynamics~\cite{gieraltowski2012multiscale,
ivanov1999multifractality}, and cloud structure~\cite{ivanova2000break}, and
many other parts of physics. In condensed-matter science, most investigations
of multifractality, which manifests itself at some phase transitions, employ a
combination of theoretical and numerical
techniques~\cite{sutradhar2019transport,
benenti2001quantum,casati2000fractal,facchini2007multifractal,jack2021visualizing,barrios2014electron, PhysRevB.102.115105, PhysRevE.104.054129}.
The experimental characterizations of multifractality in such condensed-matter
settings require high-precision experiments in good-quality samples, often at
low temperatures and at high magnetic fields. Two recent examples of such
measurements are the study of multifractal conductance fluctuations at low
magnetic fields~\cite{amin2018exotic} and the study of multifractal
superconductivity in the weak-disorder regime~\cite{rubio2020visualization, PhysRevLett.127.166802}.
By using high-mobility-graphene and tuning $\nu$, we demonstrate that
conductance fluctuations display  multifractality in the vicinity of the first
IQH plateau-to-plateau transition.

	
Our electrical-transport measurements were carried out on 
hexagonal-boron-nitride (hBN) encapsulated graphene devices with 
one-dimensional ohmic contacts  (details in Supplemental Materials). 
The electrical transport measurements were
carried out in a dilution refrigerator, with a base temperature of $20$~mK, by
using low-frequency lock-in-measurement techniques in a multi-probe
configuration at a low bias current ($\leq$1~nA) to avoid Joule heating. We
focus on our data from a particular bilayer-graphene device, 1DC8. This sample
was thermally cycled multiple times; the data we present did not change
significantly after this thermal cycling.
	
In Fig.~\ref{fig:sem-rvg}(c), we present plots of the longitudinal resistance
$R$~versus the back-gate voltage $V_G$, measured at $B=0$ and different values
of $T$. With the charge-neutrality or Dirac point at $V_D
=-2.91$~V, we find that $R$  is as low as 30~$\Omega$ at $V_G\simeq30$~V. The
field-effect mobility, estimated  at $T=20$~mK, is
$\mu=1,28,000$~cm$^2$V$^{-1}$s$^{-1}$. In the inset of
Fig.~\ref{fig:sem-rvg}(c), we show magnified plots of $R$~versus $V_G$ near the
Dirac point. Close to the Dirac point ($|\Delta V_G| = |V_G-V_D| \leq 1$~V), we
observe that $R$~increases with decreasing $T$; this suggests an insulating
state. However, for $|\Delta V_G| \geq1$~V, $R$ decreases with decreasing $T$,
indicating metallic behavior. Clearly, the effect of impurity scattering is
significantly suppressed in our sample; indeed, this is a precondition for
observing the number-density-induced insulator-metal transition in
graphene~\cite{ponomarenko2011tunable}.

Figure~\ref{fig:sem-rvg}(d) shows plots versus $\nu (= nh/eB)$ of $G_{XY}$, in
units of $e^2/h$, from our measurements at different temperatures $T$ and at
$B=16$~T. If we focus on the plot for $T$=20~mK, we observe well-developed
plateaux in $G_{XY}$ at all integer multiples of $e^2/h$, indicating a complete
lifting of the layer, spin, and valley degeneracies of the Landau-level
spectra~\cite{sarma2011electronic,kou2014electron}.  For any one of the
transitions between two adjacent plateaux, the plots of $G_{XY}$ versus $\nu$,
measured at different temperatures, intersect at a point, in the ($\nu$,
$G_{XY}$) plane. We identify each such intersection as a critical point:
($\nu_{\scriptscriptstyle  Ci}$, $G_{XY_{\scriptscriptstyle  Ci}}$), with $i$ a
positive integer, is the critical point for the $i \rightarrow (i+1)$
plateau-to-plateau quantum phase transitions from localized to delocalized
states~\cite{li2009scaling}.

	\begin {figure}[!t] \begin{center} \includegraphics[width=0.75\columnwidth,
		keepaspectratio]{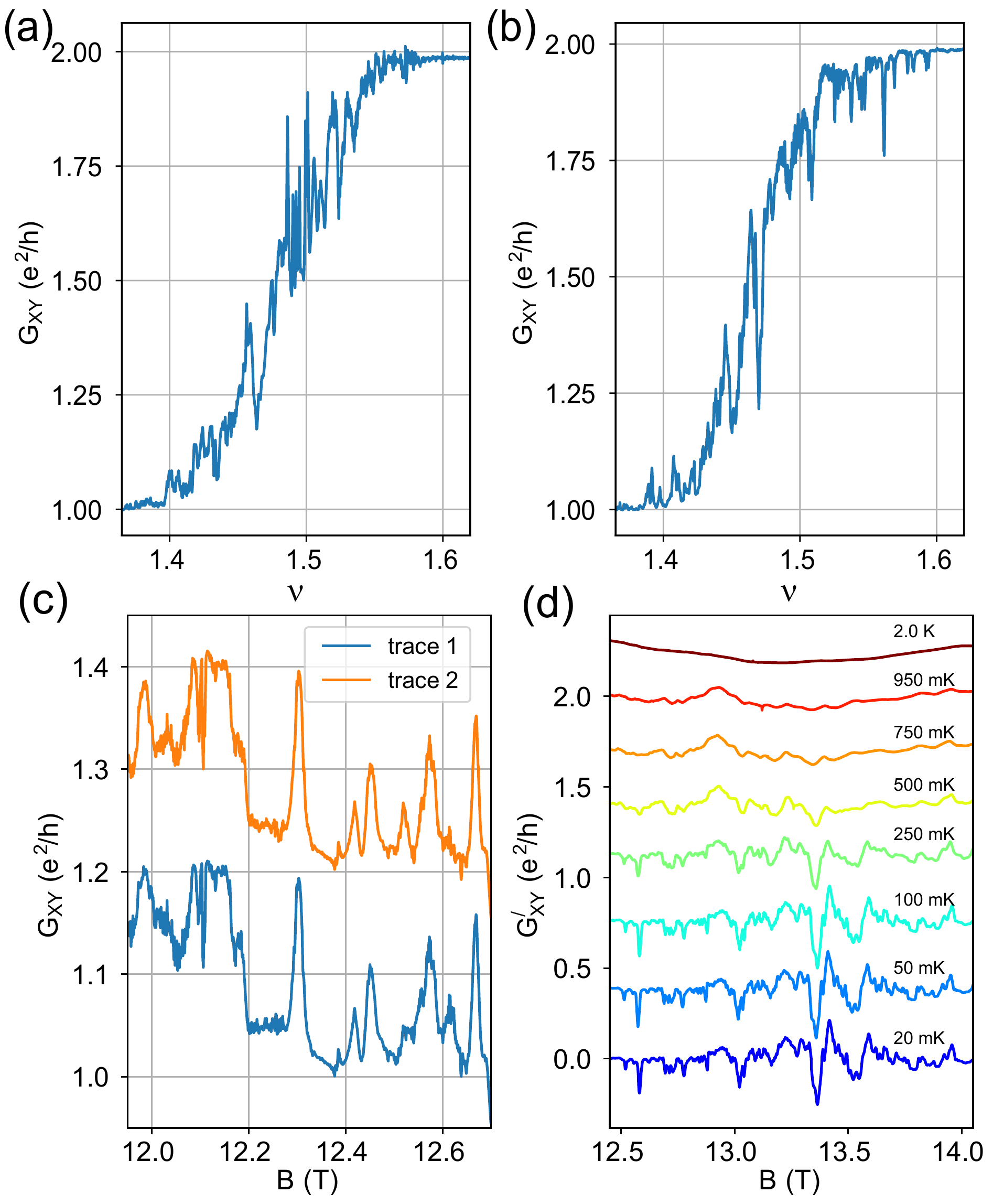} 
\caption{ Plots of $G_{XY}$ versus $\nu$
measured during the transition from  $\nu=1$ to $\nu=2$ at (a) 
fixed value of the magnetic field $B=16$~T and (b) a fixed number 
density $n = 4.75\times 10^{11} ~\mathrm{cm^{-2}} $.  (c) Plots of 
two different traces of $G_{XY}$
versus $B$, measured between $\nu=1.64$ and $\nu=1.54$, showing the
reproducibility of the mesoscopic conductance fluctuations. These 
data sets have been shifted vertically for clarity. (d) Plots of 
segments of data of $G^\prime_{XY}$ (Eq.~\ref{eqn6:bgremove}) versus 
$B$, for the $\nu=1 \longleftrightarrow \nu=2$ transition, measured at 
different temperatures $T$ and at a fixed carrier density
$n = 4.75\times 10^{11} ~\mathrm{cm^{-2}} $. Reproducible oscillations
are observed over a wide temperature ranger; the amplitude of
these oscillations decreases with increasing $T$. The data sets
have been shifted vertically for clarity. The data presented in (c) 
are from a different cool-down cycle than those in 
(a), (b), and (d).  \label{fig:pptfluct}}
	\end{center}
	
	\end {figure}
	
We now focus on the mesoscopic conductance fluctuations in the vicinity of
these critical points. In  Figs.~\ref{fig:pptfluct}(a-b), we show plots of
$G_{XY}$ versus $\nu$, for the $\nu=1 \longleftrightarrow \nu=2$ quantum-Hall
transition, measured at $T=20$~mK.  We tune $\nu$ either by changing $n$,  at
$B=16$~T (Fig.~\ref{fig:pptfluct}(a)), or by changing $B$, at $n = 4.75\times
10^{11} ~\mathrm{cm^{-2}} $ (Fig.~\ref{fig:pptfluct}(b)). These plots show that
$G_{XY}$ has significant fluctuations across the plateau-to-plateau transition.
In Fig.~\ref{fig:pptfluct}(c), we present plots of two traces of $G_{XY}$
measured at $T=20$~mK -- these data sets have been shifted vertically for
clarity. The fluctuation profiles are the same in the two traces; this
establishes that they are mesoscopic fluctuations with a unique
magnetofingerprint.  These fluctuations remain reproducible over a particular
thermal cycle; however, the detailed profile changes if we thermally cycle the
device to $T>10$~K and back.	
	
We obtain the conductance fluctuations $G_\square^\prime(x)$ from these 
measurements by subtracting a smooth background from the measured data
as follows: 
	\begin{equation}
		G_\square^\prime(x) = G_\square(x) - F[G_\square(x)].
		\label{eqn6:bgremove}
	\end{equation}
In Eq.~\ref{eqn6:bgremove}, $x$ can be $B$ or $V_G$, $G_\square$ stands for
$G_{XX}$ or $G_{XY}$, and the function $F[G_\square(x)]$ is the smooth
background in $G_\square(x)$ (see the Supplemental Material).  In
Fig.~\ref{fig:pptfluct}(d), we show representative plots of $G^\prime_{XY}$
versus $B$, from our measurements at the plateau-to-plateau transition $\nu=1
\longleftrightarrow \nu=2$, for different values of $T$ and at a fixed value of
$n = 4.75\times 10^{11} ~\mathrm{cm^{-2}} $. As we increase $T$, the mean
amplitude of the fluctuations in $G^\prime_{XY}$ decreases, but the plots of
$G^\prime_{XY}$ retain their principal features because of the
magnetofingerprint of mesoscopic fluctuations; these features go away finally
below the measurement-noise level for $T > 1$~K. Although these fluctuations
in $G^\prime_{XY}$ disappear for $T>1$~K, the plateaux of $G_{XY}$, at $e^2/h$
and $2e^2/h$, survive until much higher temperatures
[Fig.~\ref{fig:sem-rvg}(b)]. Thus, the disappearance of these conductance
fluctuations is not a consequence of the disappearance of the quantum-Hall
effect because of thermally induced level broadening.  
	
\begin {figure}[t]
	\begin{center}
		\includegraphics[width=\columnwidth,  keepaspectratio]{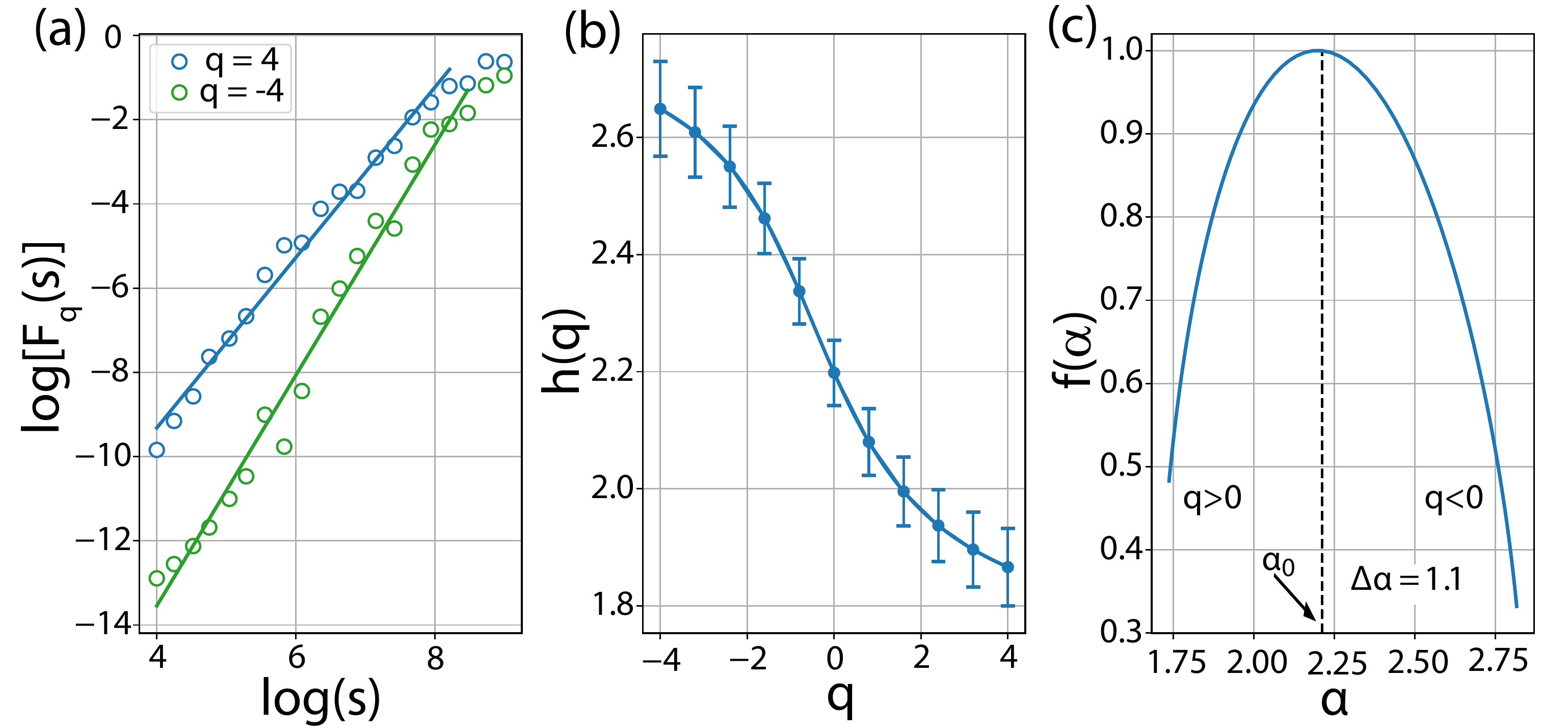}
		
\caption{(a) Plots of $\log[F_q(s)]$ versus $\log[s]$ (see Eq.~\ref{eqn:fq}),
for $q=-4$ (blue circles) and $q=4$ (green circles) for a
typical data-segment of $G^\prime_{XY}$. The thick lines are
the linear fits to the data points. (b) Plot of $h(q)$ (see
Eq.~\ref{eqn:fq}) versus $q$ for the data-segment shown in (a).
(c) Plot of the singularity spectrum $f(\alpha)$ versus
$\alpha$ (see Eq.~\ref{eqn:falpha}) that we obtain from (b).
The maximum value of this spectrum is $f_{max}(\alpha)=1$; this
is located at $\alpha_0 \simeq 2.21$ (marked by an arrow).
These data were obtained at  $\nu=1.47$ and $20$~mK.
\label{fig:mf_spectrum}}
		
	\end{center}
	\end {figure}

Having established the mesoscopic origin of the fluctuations in the conductance
across the plateau-to-plateau transitions, we now analyze the multiscaling
behavior and statistics of the fluctuations in the vicinity of the $\nu=1
\longleftrightarrow \nu=2$ critical point. Our multifractal analysis of these
fluctuations is akin to the analysis in our low-field study of universal
conductance fluctuations in single-layer graphene~\cite{amin2018exotic} (for
details of the analysis, see the Supplementary Material). Briefly, we divide
the $G^\prime_{XY}$ data-series into several segments, each centered at
different values of the filling factor $\nu$, and we compute the multifractal
spectrum as follows. We detrend each such segment and sub-divide it into $N_s$
overlapping segments, indexed by $j$, and containing $s$ data points, with
$1\leq j\leq N_s$. We obtain the generalized Hurst exponents $h(q)$ from the
power-law-scaling behavior of the order-$q$ moment of the fluctuations $F_q(s)$
by using the following relations:

\begin{eqnarray}
		g_{rms}(j) &=& \Bigg[\frac{1}{s} \sum_{i=1}^s (g_i)^2 \Bigg]^{1/2}\label{eqn:grms}; \\
		F_q(s) &=& \bigg[ \frac{1}{N_s} \sum_{j=1}^{N_s} g_{rms}(j)^q \bigg]^{1/q} \sim s^{h(q)}.
		\label{eqn:fq} 
	\end{eqnarray}
We obtain $h(q)$ for a range of values of $q$. A $q$-dependent $h(q)$ indicates
multifractality. As an example, in Fig.~\ref{fig:mf_spectrum}(a), we show
representative plots of $\log[F_q(s)]$ versus $\log[s]$, from one of our
data-series for $G^\prime_{XY}$, for $q = \pm 4$; the  circles represent the
data-points, and the thick lines are linear fits to the data.  The difference
in the slopes of the plots  suggest that $h(-4) > h(4)$; this is borne
out by the plot of $h(q)$ in Fig.~\ref{fig:mf_spectrum}(b), 
for $-4 \leq q \leq 4$.
	
Multifractality can be represented  by the singularity spectrum, which is a
plot of $f(\alpha)$ versus $\alpha$; this is obtained by the Legendre
transformation of $h(q)$ as follows:
	
\begin{eqnarray}
\begin{split}
\alpha &=& h(q) + q h^\prime(q);\\       
f(\alpha) &=& q [\alpha - h(q) ]+1	
\end{split}
\label{eqn:falpha} 
\end{eqnarray} 
	
In Fig.~\ref{fig:mf_spectrum}(c), we show a plot of $f(\alpha)$, obtained at
$20$~mK near the $\nu=1 \longleftrightarrow \nu=2$ critical point.    The width
of $f(\alpha)$, in Fig.~\ref{fig:mf_spectrum}(c), is $\Delta \alpha \simeq
1.1$.  This indicates significant multifractality of the conductance
fluctuations at this plateau-to-plateau transition.  The maximum of $f(\alpha)$
is located at $\alpha_0=2.21$ (marked by an arrow in the figure), with
$f(\alpha_0)=1$. The maximum of $f(\alpha)$ provides the support dimension, of
the data series, which is one here. In the  Supplementary Material, we show the
standard deviations of the small-amplitude fluctuations that we analyze; they
are at least ten times more than the noise level measured at the $\nu=1$
plateau.  Hence, our plot of $f(\alpha)$ is not contaminated significantly by
measurement noise.
	
We note that $f(\alpha_0)$ is asymmetrical around $\alpha_0$. To understand the
origin of this asymmetry, recall that, in the Legendre transformation
[Eq.~\ref{eqn:falpha}], the regions $q > 0$ and $q < 0$ map, respectively, onto
the $\alpha < \alpha_0$ and $\alpha > \alpha_0$ regions of the spectrum. Hence,
because of the summation procedure involved in computing $h(q)$
[Eq.~\ref{eqn:fq}], small-amplitude fluctuations in $G^\prime_{XY}$ dominate
the  $\alpha > \alpha_0$ part of $f(\alpha)$, whereas large-amplitude
fluctuations in $G^\prime_{XY}$ dominate the $\alpha < \alpha_0$ part. This
asymmetry of $f(\alpha)$ suggests, therefore, a difference between the
correlations for small- and large-amplitude fluctuations.  
	
\begin {figure}[!t]
\begin{center}
	\includegraphics[width=0.75\columnwidth, keepaspectratio]{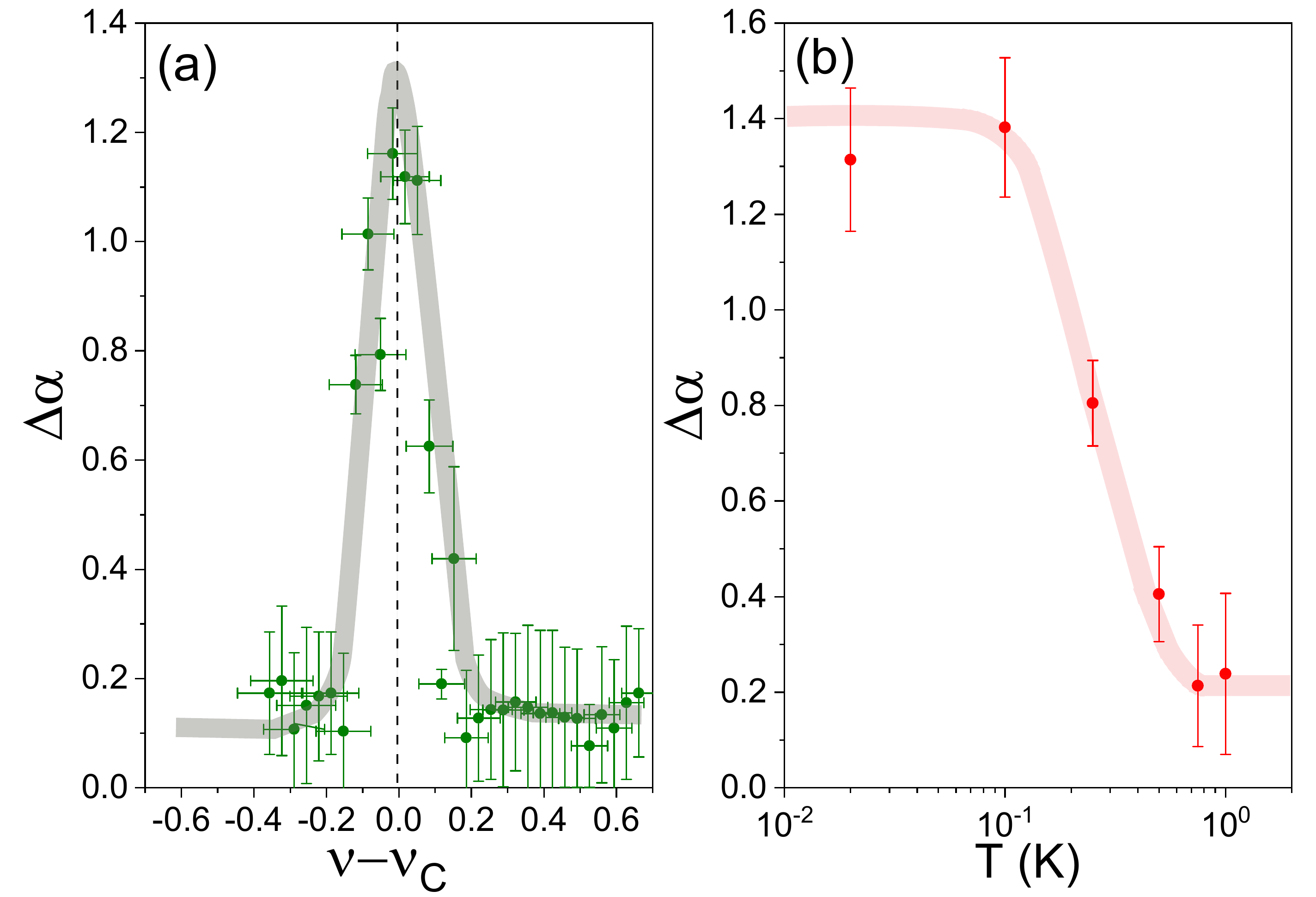}
		\caption{(a) Plot of the spectral width $\Delta \alpha$ versus
		$\nu-\nu_{\scriptscriptstyle C}$ for the device B15D4. The
		vertical line marks the $\nu=1 \longleftrightarrow \nu=2$
		critical point. The gray-shaded curve is a guide to the eye.
		The error bars for the $\nu-\nu_C$-axis are calculated from the
		the intrinsic impurity level in the device; the error bars in
		the $\Delta\alpha$ axis are derived from the error in
		calculating the slopes of the plots of $\log[F_q(s)]$ versus
		$\log[s]$. (b) Plot of $\Delta \alpha$ versus $T$ for the device
		1DC8, at a fixed value $\nu -\nu_c = 0.1$ across the 
		$\nu=1 \longleftrightarrow \nu=2$ plateau-to-plateau transition.
		The red-shaded curve is a guide to the eye.
		\label{fig:mf_nu}}
\end{center}
\end {figure}

We have also obtained the singularity spectra from plots of $G_{XY}$ versus
$V_G$ at $B=16.0$~T and $T=20$~mK in the vicinity of the $\nu=1
\longleftrightarrow \nu=2$ transition (cf. Fig. S1(b)).  We obtain the spectral
width $\Delta \alpha$ and plot it in Fig.~\ref{fig:mf_nu}(a) versus
$\nu-\nu_{\scriptscriptstyle C}$.  At $|\nu-\nu_{\scriptscriptstyle C}| =0 $,
$\Delta \alpha \simeq 1.3$, which decreases sharply to $\Delta \alpha \sim 0.2$
at $|\nu-\nu_{\scriptscriptstyle C}| \sim 0.15$. The peak in $\Delta \alpha$ at
$\nu \simeq \nu_{\scriptscriptstyle C}$ implies that the multifractality in
$G_{XY}$ increases sharply near the $\nu=1 \longleftrightarrow \nu=2$
plateau-to-plateau critical point. The maximum of $\Delta \alpha$ lies within
the error bars of our determination of $\nu_C$ from the crossing points in
Fig.~\ref{fig:sem-rvg}(d). 
	

The plot of $\Delta \alpha$ versus $\nu$ has a small but finite width away from
$\nu=\nu_{\scriptscriptstyle C}$; this indicates that $G_{XY}$ displays
multifractality not only at this IQH critical point but also in a region around
$\nu=\nu_{\scriptscriptstyle C}$. We note that the critical states are confined
to $E=E_C$ only in the thermodynamic limit. For a finite system, all states
with localization length $\xi$ larger than the system size appear to be
extended, and the distribution of physical observables (including conductance
fluctuations) is multifractal~\cite{janssen1994multifractal}. The divergence of
$\xi$, away from $\nu=\nu_C$, is suppressed only algebraically, and is governed
by $\gamma$; so the critical states, for a finite-sized system can be observed
away from $\nu_C$, with algebraically reduced probability.  Thus, our
observation of a finite amount of multifractality away from critical point
$\nu_{\scriptscriptstyle C}$ can be attributed to finite-size effects.
	
In Fig.~\ref{fig:mf_nu}(b) we plot $\Delta \alpha$ versus $T$. We note that
$\Delta \alpha$ reduces from $\simeq 1.4$ to less than $0.2$ as $T$ increases
from $100$~mK to $1.0$~K. To understand this $T$ dependence of $\Delta \alpha$,
note that the effects of quantum-interference-induced
localization-delocalization is most prominent at low temperatures.  As the $T$
increases, decoherence induced by inelastic thermal scattering reduces quantum
interference and results in delocalization~\cite{polyakov1996quantum}.  Thus,
the multifractality we observe at the $\nu=1 \longleftrightarrow \nu=2$
transition is expected to disappear with increasing $T$. As we have mentioned
earlier, this multifractality of the fluctuations in $G_{XY}$ decreases well
before the IQH plateaux disappear.
	
Earlier studies on high-mobility GaAs/AlGaAs heterostructures mesoscopic
samples have shown that the integer quantum Hall transitions are accompanied by
large, reproducible fluctuations in both $G_{XX}$ and $G_{XY}$ as functions of
$B$ and $n$~\cite{cobden1999fluctuations,feldman2012unconventional,
ilani2004microscopic}. The amplitudes of these fluctuations grow as the size of
the sample is reduced. Despite an expectation that multifractal analysis of
these fluctuations is essential for a complete description of the criticality
in the  IQH regime ~\cite{huckestein1994relation, huckestein1995scaling}, in
particular, and for topological phase transitions in general, experimental
confirmation of this multifractality has been missing hitherto.

We have presented the first experimental evidence for the multifractality of a
transport property at a topological phase transition~\cite{avron2003topological,thouless1982quantized,bellissard1994noncommutative,avron1994charge}.  Our
combination of multifractal analysis with state-of-the-art transport
measurements at a topological phase transition provide a novel method for
probing topological phase transitions in mesoscopic devices.  And our study
resolves an outstanding question in nanoscale devices, namely, the
multifractality of conductance fluctuations at such transitions in a
high-mobility 2DEG.
In particular, we have shown that conductance fluctuations display
multifractality at the integer-quantum-Hall $\nu=1 \longleftrightarrow \nu=2$
plateau-to-plateau transition in a high-mobility mesoscopic graphene device.
At this transition, we have demonstrated reproducible mesoscopic fluctuations
in $G_{XY}$ (see the Supplemental Material), with clear
multifractal spectra. This multifractality gets rapidly suppressed as $\nu$
moves away from $\nu_C$ or as $T$ is increased. We have established that, to
observe this  multifractality, it is crucial to work with very high-mobility
devices, with a well-defined critical point. 
Our results show that the multiscaling of conductance fluctuations provide a
new and clear signature for the IQH $\nu=1 \longleftrightarrow \nu=2$
plateau-to-plateau transition. Although theoretical studies have shown the
multifractality of eigenfunctions at this transition (see, e.g.,
Ref.~\cite{huckestein1995scaling,janssen1994multifractal,
peled2003observation, peled2003near, cobden1999fluctuations,
cobden1996measurement}), there has been no study hitherto of the
multifractality of transport coefficients here.  We conjecture that similar
multifractality of conductance fluctuations should also be present in (a) all
IQH plateau-to-plateau transitions, (b) the fractional-Quantum-Hall
transitions, and (c) in single-layer graphene devices. Our preliminary results
support the conjectures (a) and (c).

\begin{acknowledgments}
We than S.S. Ray for fruitful discussions. AB acknowledges
funding from DST (DST/SJF/PSA-01/2016-17). KRA  thanks CSIR, MHRD,
Govt. of India for financial support. RN thanks MHRD, Govt. of India,
for financial support. The authors thank NNfC, CeNSE, IISc for the
device fabrication facilities and MNCF, CeNSE, IISc for the device
characterization facilities. RP acknowledges support from CSIR, SERB,
and the National Supercomputing Mission (India). 
\end{acknowledgments}
	

\appendix
\setcounter{figure}{0}
\renewcommand\thefigure{S\arabic{figure}} 

\newpage

\section*{Supplementary Materials}

		\section{Characteristics of the device B15D4}
	
\begin {figure*}[h]
	\begin{center}
		\includegraphics[width=.8\textwidth, keepaspectratio]{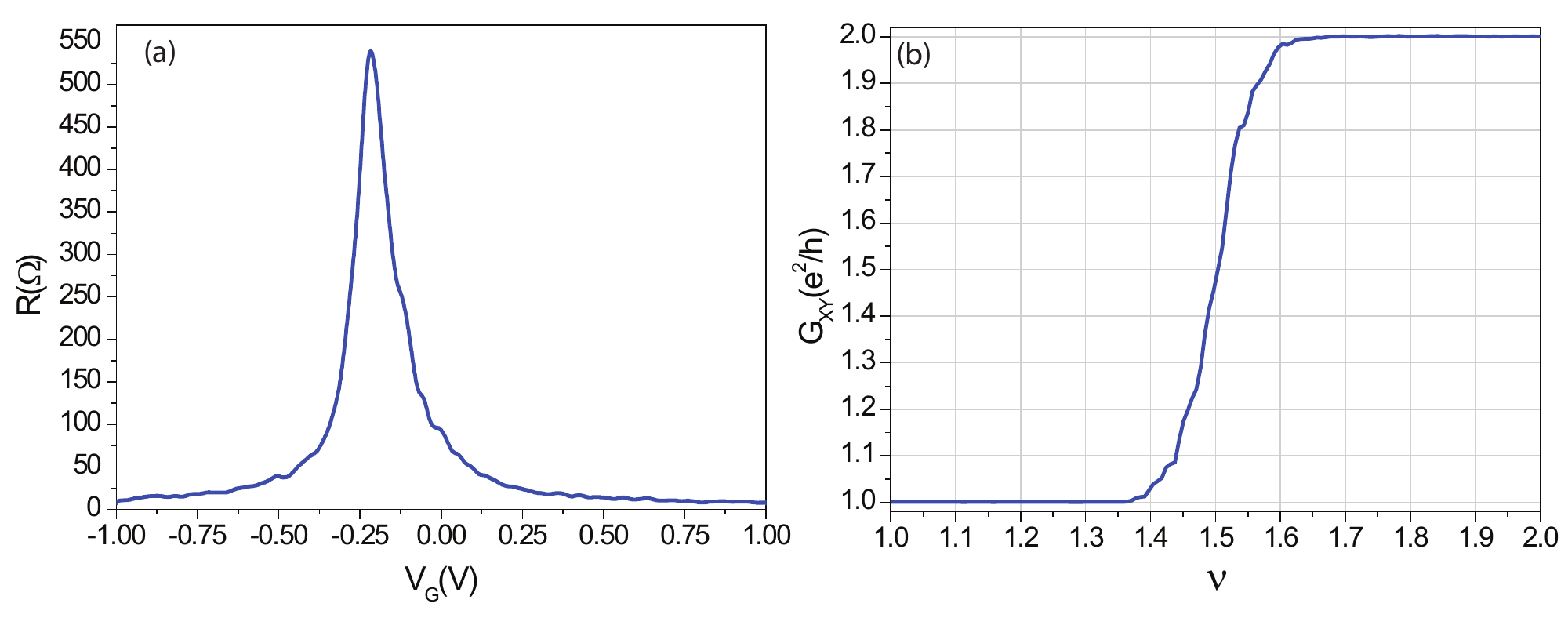}
		\caption{(a) Plot of R vs $V_G$ measured at T=20mK, in Device B15D4. (b) Plot of the quantum Hall $\nu=1$ to $\nu=2$ transition obtained at T=20mK, B=16T -- $\nu$ is tuned by sweeping the gate voltage.   
			\label{fig:device_char} } 
	\end{center}
	\end {figure*}

	In Fig.~\ref{fig:device_char}, we present the characteristics of the device B15D4. Fig.~\ref{fig:device_char}(a) shows a plot of the resistance measured as a function of the gate voltage at $T=20$~mK. The Dirac point is located at $V_D=-0.22$~V. Fig.~\ref{fig:device_char}(b) shows the quantum Hall plateau $\nu=1$ to $\nu=2$ transition obtained by sweeping the gate voltage at $T=20$~mK and $B=16$~T. 
	
	\section{Analysis of multifractality}
	
\begin {figure*}[!t]
	\begin{center}
		\includegraphics[width=.8\textwidth, keepaspectratio]{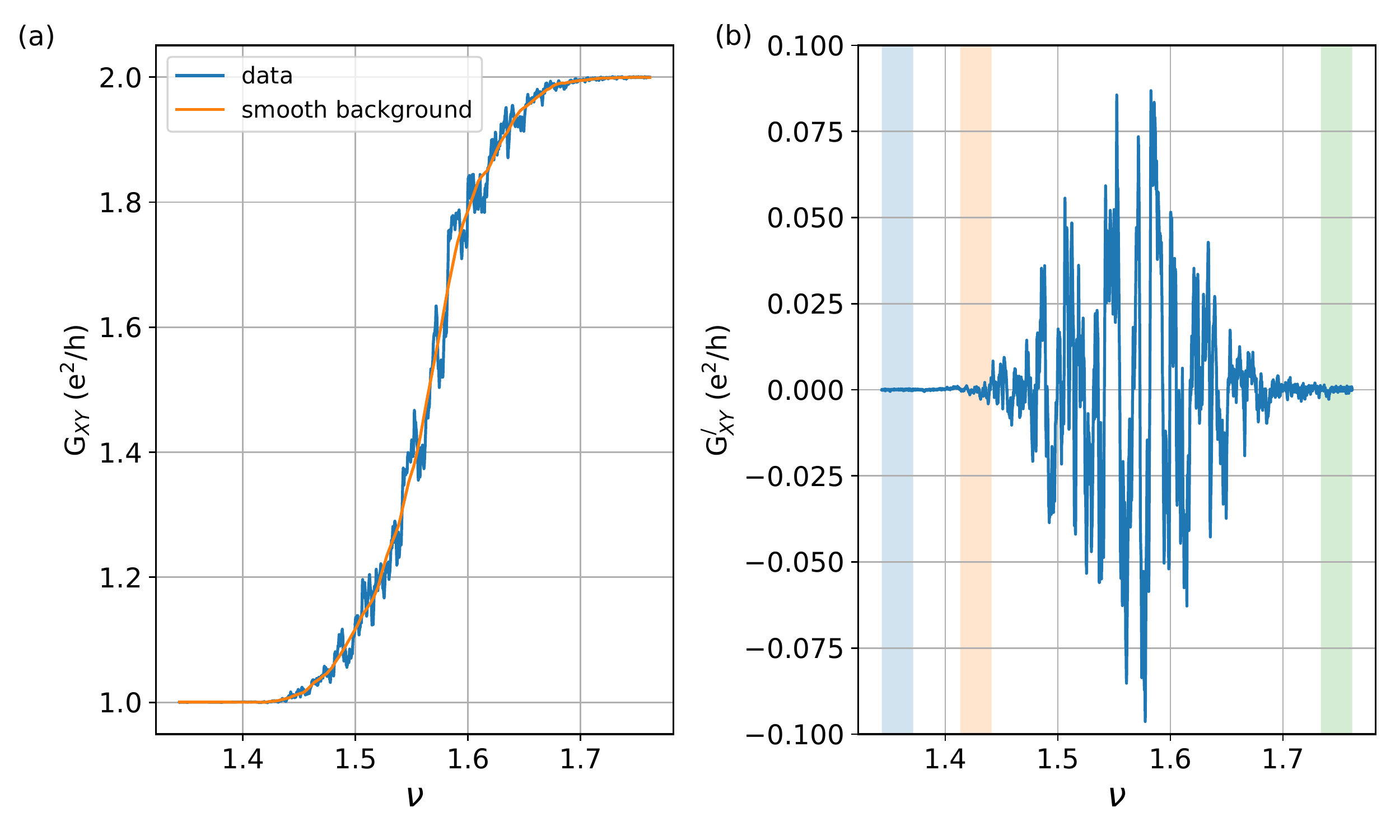}
		\caption{(a) Representative plot of  $G_{XY}$, in units of $e^2/h$ versus $\nu$ for the quantum Hall plateaus $\nu=1$  to $\nu=2$ transition is shown. The blue line is the data, while the orange line is a smooth background,  used to obtain the fluctuations in $G_{XY}$.  (b) Plot of $G^\prime_{XY}$ versus $\nu$  obtained after subtracting the background. Standard deviation, in units of  $e^2/h$, of the data were compared in the shaded regions - $1.24\times10^{-4}$ and $5.66\times10^{-4}$ obtained in blue and green shaded regions give an estimate of measurement noise level at plateau corresponding to $\nu=1$ and $\nu=2$ respectively. In contrast, the value $1.15\times10^{-3}$ observed at the orange shaded region suggests that the fluctuations that we analyze are much larger than the measurement noise level. We also note that the orange region marks onset of the multifractal fluctuations.  
			\label{fig:suppl_fluct} } 
	\end{center}
	\end {figure*}

	We explain the analysis of the conductance fluctuations to obtain the multifractal spectrum in this supplementary material. Fig.~\ref{fig:suppl_fluct}(a) shows the plot of conductance $G_{XY}$ versus $\nu$ for the quantum Hall plateaus $\nu=1$ to $\nu=2$ transition (blue line). The fluctuations observed have been confirmed to be mesoscopic fluctuations, as mentioned in the main paper. The background is marked by the solid orange line, obtained by averaging over the data. On subtracting the background, we obtain the conductance fluctuations $G'_{XY}$ as shown in Fig.~\ref{fig:suppl_fluct}(b).

	We divide $G'_{XY}$ versus $\nu$ data into overlapping segments, centered at different $\nu$  with a  fixed span in $\nu$ for further analysis. This center value of $\nu$ for each such segment is used to label the multifractal spectrum of that given data set. To give an example, the multifractal spectrum in Fig.~3(b-c) in the main text is the spectrum for such a dataset of $G'_{XY}$ versus $\nu$ centered at $\nu=1.47$. We now explain, in brief, the method to obtain multifractal singularity spectrum for each such dataset.
	
	We carry out the data analysis using the multifractal detrended fluctuation analysis (MFDFA) method~\cite{meyers2011mathematics}.
	
	\begin{enumerate}
		
		\item The dataset is divided into $N_s$ overlapping segments (indexed by j) with $s$ data points each; \{$g_i$\}, $i=1,2,...,s$.
		
		\item From each $N_s$ segment, local trend is removed by fitting a polynomial to the data. We have used a polynomial of order 1 to treat our data. Then we obtain the variance for each such segment:
		\begin{equation}
		g_{rms}(j)=\Bigg[\frac{1}{s} \sum_{i=1}^s (g_i)^2 \Bigg]^{1/2}.
		\end{equation}
		
		\item The order-$q$ moment of the fluctuations $F_q(s)$ is obtained:
		\begin{equation}
		F_q(s) = \bigg[ \frac{1}{N_s} \sum_{j=1}^{N_s} g_{rms}(j)^q \bigg]^{1/q}. 
		\end{equation}

		\item The order-$q$ moment of the fluctuations $F_q(s)$ scales with the segment size $s$ with a power  $h(q)$ as shown in  Eqn.~3 of the main paper, and is known as the generalized Hurst exponent.
		\begin{equation}
		F_q(s)  \sim s^{h(q)}.
		\end{equation}
		The scaling exponent is obtained from the slope of $\log(F_q(s))$ versus $\log(s)$ plot,  for each values of $q$, and we obtain $h(q)$.

	\end{enumerate}	
	
	\noindent The spectrum $h(q)$ versus $q$ characterizes the multifractality of a data series. Multifractality is conveniently represented using the singularity spectrum $f(\alpha)$ versus $\alpha$ which is defined as: 
	\begin {equation*}
	\tau(q)=h(q)q-1
\end{equation*}
The singularity spectrum is related to $\tau(q)$ via a Legendre transform: 
\begin {eqnarray*}
\alpha&=&\frac{d \tau(q)}{ dq} \\
f(\alpha)&=&  q\alpha - \tau(q).
\end{eqnarray*}
The spectral width $\Delta \alpha$ quantifies the multifractality of the data.

\section{Multifractality at different magnetic fields}

\begin {figure*}[!t]
\begin{center}
\includegraphics[width=0.5\textwidth, keepaspectratio]{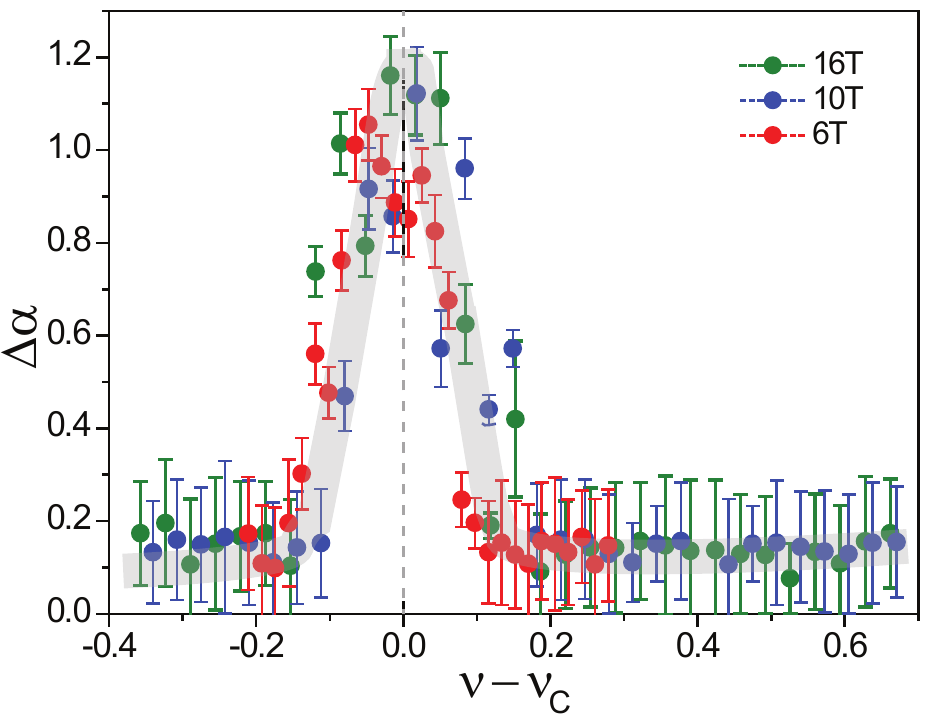}
\caption{ Plot of $\Delta \alpha$ versus $\nu-\nu_{\scriptscriptstyle C}$, where $\nu_{\scriptscriptstyle C}$ is the critical point for $\nu=1$ to $\nu=2$ transition. Data were measured at $B=16.0$~T (green filled circles), $B=10.0$~T (blue filled circles) and $B=6.0$~T (red filled circles).
\label{fig:spect_width_nu} }
\end{center}
\end {figure*}

Fig.~\ref{fig:spect_width_nu} is a plot of the multifractal spectral width $\Delta \alpha$ versus $\nu$, measured at different magnetic fields $B=16.0$~T, $B=10.0$~T, and $B=6.0$~T. Having different magnetic fields at a fixed $\nu$ corresponds to different charge carrier densities ($\nu = nh/eB$). We observe that there is no visible effect of the charge carrier density or the magnetic field  on the multifractal spectral width -- $\Delta \alpha$ depends only on $\nu$.

%

\end{document}